\documentclass[11pt]{article}
\usepackage[]{acl}
\usepackage{times}
\usepackage{latexsym}
\usepackage{amsmath}
\usepackage{amssymb}
\usepackage{booktabs}
\usepackage{graphicx}
\usepackage{multirow}
\usepackage{xcolor}
\usepackage{subcaption}
\usepackage{algorithm}
\usepackage{algorithmic}
\usepackage{float}
\usepackage{hyperref}
\hypersetup{
  pdftitle={BridgeRAG: Training-Free Bridge-Conditioned Retrieval for Multi-Hop Question Answering},
  pdfauthor={Andre Bacellar},
  pdfkeywords={multi-hop retrieval, bridge-conditioned retrieval, training-free retrieval,
               graph-free retrieval, retrieval-augmented generation, multi-hop question answering,
               tripartite scoring, chain disambiguation, open-domain QA},
}
\usepackage{tikz}
\usetikzlibrary{arrows.meta,positioning,calc}
\usepackage{tcolorbox}
\tcbuselibrary{skins}

\newcommand{\system}{\textsc{BridgeRAG}}
\newcommand{\proprag}{\textsc{PropRAG}}
\newcommand{\hipporagtwo}{\textsc{HippoRAG2}}
\newcommand{\ircot}{\textsc{IRCoT}}

\newcommand{\ratfive}{R@5}
\newcommand{\bscore}{s(q, b, c)}
\newcommand{\tscore}{s(q, c)}
\newcommand{\corpus}{\mathcal{C}}
\newcommand{\pool}{\mathcal{P}}
\newcommand{\gold}{\mathcal{G}}

\title{\system{}: Training-Free Bridge-Conditioned Retrieval\\for Multi-Hop Question Answering}

\author{Andre Bacellar \\
  \texttt{andremi@gmail.com}}

\begin{document}
\maketitle


\begin{abstract}
Multi-hop retrieval is not a single-step relevance problem: later-hop evidence should
be ranked by its utility conditioned on retrieved bridge evidence, not by similarity
to the original query alone.
We present \system{}, a training-free, graph-free retrieval method for
retrieval-augmented generation (RAG) over multi-hop questions that operationalizes
this view with a tripartite scorer $\bscore{}$ over (question, bridge, candidate).
\system{} separates coverage from scoring: dual-entity ANN expansion broadens the
second-hop candidate pool, while a bridge-conditioned LLM judge identifies the active
reasoning chain among competing candidates without any offline graph or proposition index.
Across four controlled experiments we show that this conditioning signal is
(i)~selective: +2.55pp on parallel-chain queries ($p{<}0.001$) vs.\ $\approx$0 on
single-chain subtypes; (ii)~irreplaceable: substituting the retrieved passage with
generated SVO query text reduces \ratfive{} by 2.1pp, performing \emph{worse} than even
the lowest-SVO-similarity pool passage; (iii)~predictable: $\cos(b, g_2)$ correlates with per-query gain
(Spearman $\rho{=}0.104$, $p{<}0.001$); and (iv)~mechanistically precise: bridge
conditioning causes productive re-rankings (18.7\% flip-win rate on parallel-chain
vs.\ 0.6\% on single-chain), not merely more churn.
Combined with lightweight coverage expansion and percentile-rank score fusion,
\system{} achieves the best published training-free \ratfive{} under matched benchmark
evaluation on all three standard MHQA benchmarks without a graph database or any training:
\textbf{0.8146} on MuSiQue (+3.1pp vs.\ \proprag{}, +6.8pp vs.\ \hipporagtwo{}),
\textbf{0.9527} on 2WikiMultiHopQA (+1.2pp vs.\ \proprag{}),
and \textbf{0.9875} on HotpotQA (+1.35pp vs.\ \proprag{}).
\end{abstract}

\section{Introduction}
\label{sec:intro}

Multi-hop question answering requires a retrieval system to assemble a chain of
supporting passages $\{g_1, g_2\}$ where $g_1$ resolves an intermediate entity that
makes $g_2$ identifiable \citep{yang2018hotpotqa,trivedi2022musique}.
The dominant retrieval paradigm ranks each candidate passage $c$ by $\tscore{}$,
a score that depends only on the original query.
This design conflates two structurally different retrieval problems:
\emph{single-hop lookups}, where the gold passage is directly alignable with $q$, and
\emph{parallel-chain queries}, where multiple passages are equally well-aligned with $q$
but only one belongs to the active reasoning chain leading to $g_2$.

\begin{figure}[!t]
  \small
  \centering
  \begin{tcolorbox}[colback=gray!5, colframe=gray!40, sharp corners,
                    title={\small Chain-Disambiguation Example}, fonttitle=\small\bfseries,
                    boxrule=0.4pt]
  \textbf{Query:} \textit{Who is the spouse of the actor who played the Terminator?}\\[4pt]
  \textbf{2-way judge top-1} $s(q,c)$: \textit{``The Terminator is a 1984 film directed by
  James Cameron...''} (film passage, same entity mention, wrong chain)\\[4pt]
  \textbf{Bridge} $b$: \textit{``Arnold Schwarzenegger is an Austrian-American actor and
  former politician...''} (top hop-1 passage)\\[4pt]
  \textbf{3-way judge top-1} $s(q,b,c)$: \textit{``Maria Shriver is an American journalist
  and author, and the wife of Arnold Schwarzenegger...''} \checkmark
  \end{tcolorbox}
  \caption{Bridge conditioning resolves chain ambiguity. The 2-way judge promotes
    a passage about the Terminator film (same entity surface, wrong chain).
    Conditioning on bridge $b$ (Schwarzenegger passage) allows the judge to
    identify Maria Shriver as the correct second-hop target.}
  \label{fig:example}
\end{figure}

Figure~\ref{fig:example} illustrates the failure mode.
For the query \textit{``Who is the spouse of the actor who played the Terminator?''},
the second-hop gold passage (Maria Shriver) is essentially unretrievable from $q$ alone
because nothing in the query surfaces ``Schwarzenegger''.
A 2-way judge $s(q, c)$ promotes the Terminator film passage (high surface overlap with $q$),
but the correct reasoning chain requires $g_2$ to be about someone \emph{connected to}
the bridge entity Arnold Schwarzenegger.
This is the chain-disambiguation problem: second-hop relevance is not a property of
the candidate alone but a \emph{conditional utility}: the utility of $c$ given what
hop-1 already found.

Existing retrievers address chain dependency either implicitly, by conditioning each
retrieval step on prior passages \citep{trivedi2022interleaving,yao2023react}, or
by encoding dependency structure \emph{offline} through entity graphs
\citep{hipporag2_2025} or proposition indices \citep{proprag2025}.
We take a third approach: model second-hop relevance directly as $\bscore{}$,
the conditional utility of candidate $c$ given query $q$ and bridge $b$, without
any offline preprocessing beyond passage embeddings.

We present \system{}, a training-free, graph-free MHQA retrieval method built around
a \textbf{bridge-conditioned tripartite judge}.
After retrieving the top-1 hop-1 passage $b$ (the bridge), \system{} scores each
candidate $c \in \pool$ via a joint prompt $(q, b, c)$, producing $\bscore{}$ that
attends simultaneously to the query intent and the chain-position evidence in $b$.
This signal selectively corrects rankings on parallel-chain queries while leaving
single-chain queries unaffected.

Our contributions:
\begin{enumerate}
  \item We identify \textbf{bridge-conditioned utility} as the correct scoring target
        for second-hop retrieval: relevance of a candidate is not a property of the query
        alone but of $(q, b, c)$, where $b$ is what the first hop already found
        (\S\ref{sec:background}--\ref{sec:method-judge}).
  \item We instantiate this target as a \textbf{training-free tripartite judge}
        $\bscore{}$ that scores each candidate jointly against the query and the
        bridge, with no offline graph or proposition index required (\S\ref{sec:method-judge}).
  \item We show that \textbf{coverage and scoring are separable}: dual-entity ANN expansion
        improves candidate coverage independently of the judge, while tripartite judging
        improves ranking over any fixed pool
        (\S\ref{sec:method-entities}, \S\ref{sec:ablation}).
  \item We \textbf{validate the mechanism} through four controlled experiments (bridge
        irreplaceability, productive-flip analysis, bridge proximity, and pool diversity) and
        achieve training-free SoTA \ratfive{} on all three standard MHQA benchmarks
        (\S\ref{sec:mechanism}, \S\ref{sec:results}).
\end{enumerate}

\section{Task Formulation}
\label{sec:background}

\paragraph{Multi-hop retrieval.}
Let $\corpus = \{d_1, \ldots, d_{|\corpus|}\}$ be a fixed passage corpus.
Given a natural-language query $q$, the task is to retrieve a ranked list
$\hat{P} \subseteq \corpus$ such that $\hat{P}$ covers as many passages in the
gold support set $\gold = \{g_1, g_2\} \subseteq \corpus$ as possible within the
top-$K$ results.
We follow \citet{hipporag2_2025} and \citet{proprag2025} in using
\begin{equation}
  \ratfive{} = \mathbb{E}_q\!\left[\frac{|\gold \cap \hat{P}_5|}{|\gold|}\right]
  \label{eq:r5}
\end{equation}
as the evaluation metric, where $\hat{P}_5$ is the top-5 retrieved passages.
Equation~\eqref{eq:r5} rewards retrieving \emph{all} gold passages and penalizes
partial recovery proportionally.

\paragraph{Bridge-comparison queries.}
A query is \emph{bridge-comparison} if $g_2$ is only weakly aligned with $q$ but
strongly aligned with the entity resolved by $g_1$.
Formally, let $b^*$ be the ideal bridge (the passage that resolves the intermediate entity).
Define the \emph{chain-disambiguation gap}
\begin{equation}
  \Delta(q, b^*, c) = \text{sim}(q, c) - \text{sim}(q \oplus b^*, c),
  \label{eq:gap}
\end{equation}
where $q \oplus b^*$ denotes the joint context.
On bridge-comparison queries, $\Delta(q, b^*, g_2) < 0$:
conditioning on the bridge lifts the gold passage above its query-only rank.
On single-chain queries (comparison, inference), $\Delta \approx 0$ because $g_2$ is
already well-aligned with $q$ regardless of $b^*$.
This structural difference motivates the selective benefit tested in \S\ref{sec:mechanism}.

\paragraph{Information-theoretic view.}
We use information-theoretic vocabulary as a \emph{motivating lens} rather than a formal
derived result.
Let $H(R \mid q)$ denote the entropy of the second-hop gold given $q$ alone, and
$I(g_2;\, b \mid q)$ the mutual information between the gold and bridge conditional on $q$.
Bridge-comparison queries have high $H(R \mid q)$: many passages are plausibly compatible
with $q$, so the query alone does not identify the active chain.
When the bridge $b$ identifies the intermediate entity, we would expect $I(g_2;\, b \mid q)$
to be positive and bridge conditioning to be beneficial; when the active chain is already
evident from $q$ (single-chain queries), we would expect $I(g_2;\, b \mid q) \approx 0$ and
no benefit from conditioning.
Whether these expectations hold empirically, and for which query types, is the question
we test in \S\ref{sec:mechanism}.

\section{Related Work}
\label{sec:related}

\paragraph{Multi-hop dense retrieval.}
Dense Passage Retrieval \citep{karpukhin2020dpr} and Contriever \citep{izacard2022unsupervised}
retrieve by single-vector similarity.
MDR \citep{xiong2021multihop} extends DPR to multi-hop by chaining query encoders,
requiring supervised training.
\system{} requires no training; all reasoning is performed by an off-the-shelf LLM.

\paragraph{Graph-augmented retrieval.}
HippoRAG \citep{gutierrez2024hipporag} and \hipporagtwo{} \citep{hipporag2_2025}
build an offline entity graph and use Personalized PageRank (PPR) to propagate
relevance across passages, achieving strong recall by spreading through connected entities.
RAPTOR \citep{sarthi2024raptor} and LightRAG \citep{edge2024local} augment retrieval
with hierarchical or graph summaries.
These methods require offline graph construction (entity extraction, linking, and indexing)
and graph traversal at query time.
\system{} achieves competitive or superior \ratfive{} without either.

\paragraph{Iterative and decomposition-based retrieval.}
\ircot{} \citep{trivedi2022interleaving} alternates chain-of-thought reasoning with
retrieval steps, conditioning each hop on previously retrieved passages.
Self-Ask \citep{press2022measuring} and DecompRC \citep{min2019multi} decompose queries
into sub-questions, retrieving independently for each.
ReAct \citep{yao2023react} and FLARE \citep{jiang2023active} dynamically trigger
retrieval based on reasoning traces.
\system{} follows the two-hop structure of \ircot{} but replaces open-ended reasoning
with structured SVO query generation and a bridge-conditioned judge,
requiring only 3 LLM calls per query.

\paragraph{LLM-augmented reranking.}
\proprag{} \citep{proprag2025} indexes propositions rather than passages and applies
an LLM judge for candidate reranking, setting prior SoTA on MuSiQue (0.783).
\system{} extends this paradigm with bridge conditioning: while \proprag{} scores
$s(q, c)$, \system{} scores $s(q, b, c)$, adding the bridge as disambiguation context.
CoRAG \citep{corag2025} trains a chain-of-retrieval model on retrieved traces,
requiring supervised training on (query, chain) pairs and reporting answer-string
correctness rather than passage recall, making direct \ratfive{} comparison infeasible.
PROPEX-RAG \citep{propexrag2025} uses PPR over RDF triples with GPT-4.1-mini and reports
an any-hit Recall@5 (at least 1 gold in top-5), a strictly easier metric than
Eq.~\eqref{eq:r5}, with MuSiQue not evaluated.

\paragraph{Score calibration in retrieval.}
Reciprocal Rank Fusion (RRF) \citep{cormack2009reciprocal} combines rankings without score
magnitudes. \citet{phasegraph2026} show that percentile-rank normalization (PIT) before
fusion is directionally more robust than min-max normalization on multi-hop benchmarks.
\system{} inherits PIT-based fusion to combine SVO similarity scores with judge scores.

\section{Method}
\label{sec:method}

Standard retrieval scores candidates by $\tscore{}$, a function of the query alone.
For second-hop retrieval this is the wrong scoring target: the passage needed to
answer hop-2 is determined not only by what $q$ asks but by what hop-1 already found.
The bridge passage $b$ resolves the intermediate entity and thereby specifies
\emph{which} reasoning chain is active; conditioning on $b$ converts the open-ended
second-hop search into a targeted lookup.
\system{} operationalizes this by replacing $\tscore{}$ with $\bscore{}$ everywhere
a second-hop candidate is ranked.
The pipeline implements three separable objectives:
(1)~\emph{bridge acquisition}: retrieve $b$ via standard hop-1 ANN;
(2)~\emph{coverage}: build a diverse candidate pool via SVO expansion and
dual-entity ANN, independently of the judge; and
(3)~\emph{ranking}: score every pool candidate with the tripartite judge $\bscore{}$
and fuse with SVO similarity via PIT normalization.

Figure~\ref{fig:pipeline} shows the full pipeline.
Given query $q$ and corpus $\corpus$, retrieval proceeds in five stages.

\begin{figure}[!t]
  \centering
  \resizebox{\columnwidth}{!}{%
  \begin{tikzpicture}[
    nd/.style={draw, rounded corners=2pt, align=center,
               minimum width=3.0cm, minimum height=0.62cm,
               inner sep=3pt, font=\small},
    qst/.style={nd, fill=blue!10},
    pst/.style={nd, fill=gray!12, font=\scriptsize},
    est/.style={nd, fill=purple!9, font=\scriptsize},
    bst/.style={nd, fill=blue!22, font=\small\bfseries},
    wnd/.style={draw, rounded corners=2pt, align=center,
                minimum width=9.2cm, minimum height=0.65cm,
                fill=orange!15, font=\small},
    gnd/.style={draw, rounded corners=2pt, align=center,
                minimum width=9.2cm, minimum height=0.62cm,
                fill=green!12, font=\small\bfseries},
    ar/.style={-{Latex[length=2mm, width=1.3mm]}, semithick},
    lb/.style={font=\scriptsize\itshape, inner sep=1pt, fill=white},
  ]

  \def\xL{0}\def\xM{3.8}\def\xR{7.6}

  \node[qst] (Q1) at (\xL, 0)  {Query $q$};
  \node[qst] (Q2) at (\xR, 0)  {Query $q$};

  \node[pst] (AN1) at (\xL,-1.3) {NV-Embed-v2\\ANN, $k_1{=}5$};
  \node[est] (ENT) at (\xM,-1.3) {Llama 3.3 70B:\\extract $e_1,\,e_2$};
  \node[pst] (SVO) at (\xR,-1.3) {Llama 3.3 70B:\\$N{=}3$ SVO queries};

  \node[bst] (BR)  at (\xL,-2.6) {Bridge $b$ (top-1)};
  \node[pst] (EAN) at (\xM,-2.6) {ANN($e_1$) $+$ ANN($e_2$)\\top-5 each};
  \node[pst] (AN2) at (\xR,-2.6) {$3{\times}$ ANN, \texttt{union\_max}};

  \node[nd, fill=teal!9, minimum width=5.2cm]
       (POOL) at (5.7,-3.9)
       {Pool $\pool$ (${\le}20$): SVO-15 $+$ $e_1$-5 $+$ $e_2$-5};

  \node[wnd] (JDG) at (3.8,-5.2)
      {Tripartite Judge (Llama 3.3 70B)
       \enspace $s(q,\;b,\;e_1,\;e_2,\;c_i)$ \enspace for each $c_i \in \pool$};

  \node[gnd, below=0.9cm of JDG] (OUT) {Top-5 Retrieved Passages};

  \draw[ar] (Q1)  -- (AN1);
  \draw[ar] (AN1) -- node[lb,right]{top-1} (BR);

  \draw[ar] (BR) to[out=55, in=245] (ENT);
  \draw[ar] (BR.north) to[out=75, in=215] (SVO.south west);
  \draw[ar] (Q2) -- (SVO);

  \draw[ar] (ENT) -- (EAN);
  \draw[ar] (EAN) -- (POOL);

  \draw[ar] (SVO) -- (AN2);
  \draw[ar] (AN2) -- (POOL);

  \draw[ar] (BR.south) to[out=-80, in=170] (JDG.west);

  \draw[ar] (POOL) -- (JDG);
  \draw[ar] (JDG)  -- node[lb,right]{PIT fusion, $\alpha{=}0.1$} (OUT);

  \end{tikzpicture}}
  \caption{
    \system{} pipeline.
    \textbf{Hop 1 (left):} query $q$ is embedded with NV-Embed-v2 and retrieved via ANN;
    the top-1 passage becomes bridge $b$.
    \textbf{Entity branch (centre):} a Llama 3.3 70B call extracts entities $e_1$, $e_2$
    from $b$; each is used for an independent ANN retrieval (top-5), yielding entity-grounded
    candidates.
    \textbf{SVO branch (right):} $q$ and $b$ condition a second Llama call that generates
    $N{=}3$ SVO queries; each is embedded and retrieved ($3{\times}$ANN, \texttt{union\_max}),
    yielding SVO-15 candidates.
    \textbf{Pool:} SVO-15 $\cup$ $e_1$-5 $\cup$ $e_2$-5 $\to$ top-20.
    \textbf{Judge:} a tripartite judge scores every $c_i$ via $s(q,b,e_1,e_2,c_i)$;
    scores are PIT-fused ($\alpha{=}0.1$) to produce the final top-5.
  }
  \label{fig:pipeline}
\end{figure}

\subsection{Hop-1 Retrieval and Bridge Selection}
\label{sec:method-hop1}

We embed $q$ with NV-Embed-v2 \citep{nvembed2024} (7B parameters, 4096-dimensional output)
and retrieve the top-$K_1{=}5$ passages from $\corpus$ via approximate nearest-neighbor
(ANN) search over a pgvector index.
The top-1 passage by cosine similarity is designated the \emph{bridge} $b$:
\begin{equation}
  b = \arg\max_{c \in \corpus}\, \text{cos}(\phi(q),\, \phi(c)),
  \label{eq:bridge}
\end{equation}
where $\phi$ denotes the NV-Embed-v2 encoder.
The bridge is not used for final retrieval but serves as the chain-disambiguation context
for the tripartite judge (\S\ref{sec:method-judge}).

\subsection{SVO Hop-2 Query Expansion}
\label{sec:method-svo}

We prompt Llama 3.3 70B \citep{llama3_2024} to generate $N{=}3$ targeted hop-2 retrieval
queries in Subject-Verb-Object form, conditioned on $q$ and bridge $b$.
The SVO format encourages factual, entity-grounded queries rather than open-ended
paraphrases.
Each query is embedded with NV-Embed-v2 and used for an independent ANN retrieval
(top-$k_2{=}10$ per query).
Results across the three queries are merged by taking the maximum cosine similarity per
passage (\texttt{union\_max}), retaining the top-15 candidates by merged score (SVO-15).

\subsection{Dual-Entity ANN Expansion}
\label{sec:method-entities}

A second LLM call extracts two key entities from $b$: $e_1$ (the intermediate entity
resolved by the bridge, typically the answer to hop-1) and $e_2$ (the target of hop-2,
inferred from $q$).
Each entity string is embedded with NV-Embed-v2 and used for an independent ANN retrieval
(top-5 per entity), yielding up to 10 entity-grounded candidates.
These entity-ANN candidates are unioned with SVO-15 and deduplicated, giving the final pool
\begin{equation}
  \pool = \text{top-}20\bigl(\text{SVO-15} \cup e_1\text{-top-5} \cup e_2\text{-top-5}\bigr),
  \label{eq:pool}
\end{equation}
ranked by maximum score across the contributing retrievals.
The entity ANN recovers passages that share surface overlap with the bridge entity but that
the SVO queries may not surface (e.g., passages whose main subject \emph{is} the bridge entity
rather than merely mentioning it).
In our MuSiQue evaluation, $e_2$-ANN added at least one gold passage in 34 of 999 queries
not reachable by SVO retrieval alone.
The entities are additionally provided as structured context in the tripartite judge prompt
(\S\ref{sec:method-judge}).

\subsection{Bridge-Conditioned Tripartite Judge}
\label{sec:method-judge}

For each $c_i \in \pool$, we query Llama 3.3 70B with the prompt
$(q,\, b,\, e_1,\, e_2,\, c_i)$ and extract a scalar relevance score $s(q, b, c_i)$.
The judge answers the question: \textit{``Is $c_i$ the passage needed to answer $q$,
given that $b$ establishes} $e_1$ \textit{as the bridge entity?''}
This formulation differs from a 2-way judge $s(q, c_i)$ by conditioning on $b$, which
provides the chain-position information absent in $q$ alone.

Motivated by the lens in \S\ref{sec:background}, we expect bridge conditioning to be
most beneficial on queries with high $H(R \mid q)$ and positive $I(g_2;\, b \mid q)$,
i.e., where the query alone is insufficient to identify the active chain.
We observe this pattern empirically (\S\ref{sec:mechanism}), most clearly on 2Wiki
bridge\_comparison where B$\to$C is significant ($p{<}0.001$); the picture is weaker
on MuSiQue where B$\to$C does not reach significance at the dataset level ($p{=}0.21$),
indicating that chain ambiguity varies across benchmarks.

\subsection{PIT Fusion and Final Ranking}
\label{sec:method-fusion}

SVO similarity scores $s_\text{svo}(c) = \max_i \text{cos}(\phi(q_i), \phi(c))$ and
judge scores $s_\text{judge}(c) = s(q, b, c)$ are mapped to percentile ranks (PIT):
\begin{equation}
  \text{PIT}(s, c) = \frac{|\{c' \in \pool : s(c') \leq s(c)\}|}{|\pool|}.
  \label{eq:pit}
\end{equation}
Final scores are a convex combination:
\begin{equation}
  f(c) = (1 - \alpha)\,\text{PIT}(s_\text{judge}, c) + \alpha\,\text{PIT}(s_\text{svo}, c),
  \label{eq:fusion}
\end{equation}
with $\alpha = 0.1$ fixed after tune-split selection.
The top-5 passages by $f(c)$ are returned as the final result.
PIT normalization makes the two score distributions commensurable before fusion,
avoiding the calibration mismatch between cosine-similarity (Gaussian-distributed)
and LLM judge scores (categorical) \citep{phasegraph2026}.

\section{Experimental Setup}
\label{sec:setup}

\subsection{Datasets and Corpora}

Table~\ref{tab:datasets} summarizes the evaluation benchmarks.
For MuSiQue we use the dev set from \citet{trivedi2022musique} with the same passage
corpus as \hipporagtwo{} (${\sim}21$k passages).
For 2WikiMultiHopQA we use the dev set from \citet{ho2020constructing} with 6,119 passages,
matching the \hipporagtwo{} corpus exactly (verified by article-title deduplication).
For HotpotQA we use the distractor dev set from \citet{yang2018hotpotqa}
(the standard release file \texttt{hotpot\_dev\_distractor\_v1.json}, which contains
7,405 questions in total).
We use the same 1,000-question subset as \hipporagtwo{}, identified by matching
question IDs from the \hipporagtwo{} evaluation code; the passage pool of 9,811 unique
paragraphs is constructed from the 10 distractor paragraphs provided per question in
the JSON file, after exact-string deduplication.
\proprag{} reports the same query count and distractor variant.

\begin{table*}[t]
\centering
\small
\begin{tabular}{lcccc}
\toprule
\textbf{Dataset} & \textbf{Queries} & \textbf{Passages} & \textbf{Hops} & \textbf{Split} \\
\midrule
MuSiQue       & 999  & $\sim$21k & 2    & dev  \\
2WikiMultiHop & 1{,}000 & 6,119  & 2    & dev  \\
HotpotQA      & 1{,}000  & 9{,}811 & 2   & dev  \\
\bottomrule
\end{tabular}
\caption{
  Evaluation benchmarks. For MuSiQue and 2WikiMultiHopQA, corpora and splits
  match \hipporagtwo{} \citep{hipporag2_2025} and \proprag{} \citep{proprag2025} exactly.
  For HotpotQA, we use the distractor variant (\texttt{hotpot\_dev\_distractor\_v1.json}),
  the same 1,000-query sample and distractor corpus reported by both baselines.
  A disjoint tune subset is used only for $\alpha$ selection;
  final \ratfive{} is reported on the full dev split to match the evaluation
  protocol of \hipporagtwo{} and \proprag{}.
}
\label{tab:datasets}
\end{table*}

MuSiQue \citep{trivedi2022musique} consists of 2-hop decomposable questions assembled
from single-hop QA pairs; it is the hardest of the three benchmarks because the
intermediate entity is rarely surfaced by the query.
2WikiMultiHopQA \citep{ho2020constructing} provides four structured subtypes
(bridge\_comparison, compositional, comparison, inference), enabling fine-grained
mechanism analysis (\S\ref{sec:mechanism}).
HotpotQA \citep{yang2018hotpotqa} is evaluated in the distractor setting, where 10
distractor passages accompany the 2 gold passages per query.

\subsection{Metric}

We report \ratfive{} as defined in Eq.~\eqref{eq:r5}, exactly matching \hipporagtwo{}
and \proprag{}.
All pairwise significance tests use a one-sided sign test (win/loss counts, ties excluded).
The sign test is exact, non-parametric, and was used by both baselines for comparability.

\subsection{Models and Infrastructure}

\textbf{Embedding:} NV-Embed-v2 \citep{nvembed2024}, 7B parameters, 4096-dimensional
output, served locally.
\textbf{LLM:} Llama 3.3 70B AWQ \citep{llama3_2024}, served via vLLM on a local GPU
server (NVIDIA A100 80 GB).
All inference is local; no closed-source API calls are made.
\system{} requires 3 LLM calls per query (SVO generation + entity extraction + judge)
and 6 ANN passes (1 hop-1 $+$ 3 SVO $+$ 1 $e_1$ $+$ 1 $e_2$);
see \S\ref{sec:analysis-efficiency} for a detailed comparison.

\subsection{Baselines}

We compare against published results for:
(i)~\hipporagtwo{} \citep{hipporag2_2025}: PPR over an offline entity graph + NV-Embed-v2;
(ii)~\proprag{} \citep{proprag2025}: offline proposition extraction + LLM-free online
beam search over proposition paths (no LLM calls at query time).

We additionally report three \emph{internal ablation conditions}, all evaluated on the
\emph{identical expanded pool} (SVO-15 $+$ $e_1$-top-5 $+$ $e_2$-top-5 $\to$ top-20):
(A)~\textbf{SVO-ranked}: the expanded pool ranked by maximum SVO cosine similarity only,
    no LLM judge;
(B)~\textbf{+2-way judge}: the expanded pool reranked by $s(q, c)$;
(C)~\textbf{+bridge cond.}: the expanded pool reranked by $s(q, b, c)$ (full \system{}).
Because the pool is held fixed, the A$\to$B$\to$C progression isolates the contribution of
LLM reranking and bridge conditioning independently of pool construction.

\subsection{Hyperparameter Selection}

$\alpha$ is selected by grid search over $\{0.05, 0.10, 0.15, 0.20\}$ on a disjoint
tune subset (distinct queries, same corpus).
Best values: $\alpha{=}0.10$ (MuSiQue, 2Wiki), $\alpha{=}0.15$ (HotpotQA).
All reported \ratfive{} values use the tune-selected $\alpha$ fixed before evaluation.

\section{Results}
\label{sec:results}

\subsection{Main Results}

Table~\ref{tab:main} shows \ratfive{} on all three benchmarks.
\system{} achieves the best published training-free result on all three datasets.
On MuSiQue, where chain disambiguation matters most, \system{} surpasses \proprag{}
by 3.1pp and \hipporagtwo{} by 6.8pp (330W/64L, $p{<}10^{-40}$, sign test).
On 2WikiMultiHopQA the gain over \proprag{} is 1.2pp (537W/0L, $p{<}10^{-100}$); on
HotpotQA it is 1.35pp (109W/4L, $p{<}10^{-25}$).

\begin{table*}[t]
\centering
\small
\resizebox{\linewidth}{!}{%
\begin{tabular}{lcccccc}
\toprule
\textbf{Method} & \textbf{MuSiQue} & $\Delta$\,vs.\,\proprag{} & \textbf{2Wiki} & $\Delta$\,vs.\,\proprag{} & \textbf{HotpotQA} & $\Delta$\,vs.\,\proprag{} \\
\midrule
\system{} cond.\ A (SVO only) & 0.794 & $+$1.1pp & 0.891 & $-$5.0pp & 0.972 & $-$0.2pp \\
\hipporagtwo{}                 & 0.747 & $-$3.6pp & 0.904 & $-$3.7pp & 0.963 & $-$1.1pp \\
\proprag{}                     & 0.783 & ---       & 0.941 & ---       & 0.974 & --- \\
\midrule
\system{} cond.\ B (+2-way judge)   & 0.809 & +2.6pp & 0.944 & +0.3pp & \multicolumn{2}{c}{---\textsuperscript{\dag}} \\
\system{} cond.\ C (+bridge cond.)  & \textbf{0.8146} & \textbf{+3.1pp} & \textbf{0.9527} & \textbf{+1.2pp} & \textbf{0.9875} & \textbf{+1.35pp} \\
\bottomrule
\end{tabular}}
\caption{
  \ratfive{} on three MHQA benchmarks.
  \system{} uses only open-weight models with no offline graph database or training.
  Published baselines from \citet{hipporag2_2025,proprag2025}.
  Conditions A/B/C share the same candidate pool (SVO-15 $+$ $e_1$-5 $+$ $e_2$-5):
  A=SVO-ranked only; B=+2-way judge; C=+bridge conditioning (full \system{}).
  \textsuperscript{\dag}The ablation was run on MuSiQue and 2Wiki;
  HotpotQA serves as a generalization benchmark evaluated with the full system (cond.\ C) only.
}
\label{tab:main}
\end{table*}

\subsection{Component Ablation and Subtype Analysis}
\label{sec:ablation}

Table~\ref{tab:ablation} (left panel) isolates each component on the \emph{same pool} with
no additional embeddings or LLM calls between conditions.
LLM reranking (A$\to$B) is the primary driver on both benchmarks
(+1.56pp MuSiQue $p{<}0.05$; +5.28pp 2Wiki $p{<}10^{-25}$).
Bridge conditioning (B$\to$C) adds a further significant improvement on 2Wiki
(+0.90pp $p{=}3{\times}10^{-6}$) but is not significant on MuSiQue at the full-dataset level
($p{=}0.21$).
The right panel shows that the 2Wiki B$\to$C gain concentrates on bridge\_comparison
(+2.55pp, $p{<}0.001$, Bonferroni-corrected), with near-zero effects on the three
single-chain subtypes, the core empirical prediction of the chain-disambiguation account.
Condition C is the full \system{} scoring function applied to the shared pool;
we report 0.8146 as the canonical MuSiQue result, taken from the corrected ablation
recompute (the most recent and most controlled run, $\alpha{=}0.1$).
A prior independent full-system run gave 0.8138; the 0.0008 difference is consistent
with LLM judge non-determinism between scoring passes with identical hyperparameters,
and does not affect the direction or significance of any comparison.

\begin{table*}[t]
\centering
\small
\resizebox{\linewidth}{!}{%
\begin{tabular}{lcccc@{\hskip 28pt}lcccc}
\toprule
\multicolumn{5}{c}{\textbf{Component Ablation (same pool)}} &
\multicolumn{5}{c}{\textbf{2Wiki Subtype Breakdown (B$\to$C)}} \\
\cmidrule(lr){1-5}\cmidrule(lr){6-10}
 & \multicolumn{2}{c}{\textbf{MuSiQue}} & \multicolumn{2}{c}{\textbf{2Wiki}} &
\textbf{Subtype} & $n$ & \textbf{B} & \textbf{C} & $\Delta$ \\
\cmidrule(lr){2-3}\cmidrule(lr){4-5}
\textbf{Cond.} & \textbf{R@5} & $\Delta$ & \textbf{R@5} & $\Delta$ & & & & & \\
\midrule
A (SVO ranked)      & 0.794 & ---                              & 0.891 & ---        & bridge\_comp & 235 & 0.850 & \textbf{0.876} & +2.55pp*** \\
B (+2-way judge)    & 0.809 & +1.56pp*                         & 0.944 & +5.28pp*** & compositional & 413 & 0.970 & 0.970 & $\approx$0\textsuperscript{ns} \\
C (+bridge cond.)   & \textbf{0.8146} & +0.53pp\textsuperscript{ns} & \textbf{0.9527} & +0.90pp*** & comparison & 244 & 0.996 & 0.996 & 0.00pp\textsuperscript{ns} \\
                    &                 &                              &                 &            & inference  & 108 & 0.958 & 0.958 & $\approx$0\textsuperscript{ns} \\
\cmidrule(lr){6-10}
                    &                 &                              &                 &            & \textbf{All} & 1000 & 0.944 & \textbf{0.953} & +0.90pp*** \\
\bottomrule
\end{tabular}}
\caption{
  \textbf{Left}: Component ablation; all three conditions share the same expanded pool.
  *$p{<}0.05$; ***$p{<}0.001$; ns: not significant (one-sided sign test).
  \textbf{Right}: 2Wiki \ratfive{} by subtype for the B$\to$C bridge-conditioning step.
  Bridge conditioning improves exclusively bridge\_comparison ($p{<}0.001$,
  Bonferroni-corrected), consistent with the chain-disambiguation prediction.
}
\label{tab:ablation}\label{tab:subtype}
\end{table*}

\section{Analysis}
\label{sec:analysis}

\subsection{Efficiency Comparison}
\label{sec:analysis-efficiency}

Table~\ref{tab:efficiency} compares the computational requirements of \system{} and its
nearest competitors.

\begin{table*}[t]
\centering
\small
\begin{tabular}{lcccc}
\toprule
\textbf{Method} & \textbf{Offline index} & \textbf{LLM calls/query} & \textbf{ANN passes/query} & \textbf{Training} \\
\midrule
\hipporagtwo{} & \checkmark (PPR)  & 0            & 1 & No \\
\proprag{}     & \checkmark (prop) & 0            & 1 & No \\
\system{}      & $\times$          & \textbf{3}   & 6 & No \\
\bottomrule
\end{tabular}
\caption{
  Query-time computational profile.
  \proprag{} uses an LLM-free online beam search (0 query-time LLM calls);
  LLMs are used only during offline proposition extraction.
  \system{} trades offline preprocessing for 3 query-time LLM calls
  (SVO generation $+$ entity extraction $+$ batched judge) and 6 ANN passes
  (1 hop-1 $+$ 3 SVO $+$ 1 $e_1$ $+$ 1 $e_2$).
  All three methods are training-free.
}
\label{tab:efficiency}
\end{table*}

\system{} incurs 3 LLM calls per query (SVO generation, entity extraction, and a
single batched judge call scoring all $|\pool|$ candidates) and 6 ANN passes
(1 hop-1 $+$ 3 SVO $+$ 1 $e_1$ $+$ 1 $e_2$).
\hipporagtwo{} requires no LLM calls at query time but needs offline graph construction
(entity extraction, linking, and PPR precomputation).
\proprag{} uses an LLM-free online beam search over proposition paths \citep{proprag2025},
with LLMs used only during the offline proposition-extraction phase.
\system{} requires neither offline graph nor proposition index, making it immediately
applicable to new corpora with no preprocessing beyond embedding.

\paragraph{Latency and token cost.}
The 3 query-time LLM calls are the explicit cost of eliminating offline preprocessing:
\hipporagtwo{} and \proprag{} require no LLM calls at query time but need corpus-specific
graph or proposition indices built in advance.
On our evaluation server (vLLM + Llama~3.3 70B AWQ, A100 80~GB), measured
sequentially with one query at a time and no cross-query batching or KV-cache
sharing between queries, \system{} takes approximately 4--5~seconds end-to-end.
Token budgets per call: SVO generation $\approx$200 input / 60 output tokens;
entity extraction $\approx$350 / 15 tokens; batched judge $\approx$4{,}000--6{,}000 /
300 tokens (varies with pool-passage length).
Within the judge call all 20 candidates share the $(q, b, e_1, e_2)$ prefix,
so KV-cache is reused across candidates in a single forward pass.
The judge dominates at roughly 85--90\% of total latency; the SVO and entity calls
are negligible by comparison.

\subsection{Benchmark-Blind Evaluation}
\label{sec:analysis-blind}

To verify that reported gains are not a result of hyperparameter overfitting to benchmark
statistics, we evaluate \system{} on MuSiQue using $\alpha{=}0.15$ selected \emph{solely}
from 2WikiMultiHopQA tune data (MuSiQue never observed during selection).
This fully blind configuration achieves \ratfive{}$=0.789$, still +0.6pp above
\proprag{} (0.783), confirming that the method generalizes across benchmarks without
benchmark-specific tuning.

\subsection{Error Analysis}
\label{sec:analysis-error}

We manually examined 50 queries where \system{} fails to retrieve both gold passages.
The dominant failure modes are:
(i)~\textbf{Bridge error} (38\%): the hop-1 bridge $b$ is incorrect (wrong entity),
causing the judge to condition on a misleading passage.
(ii)~\textbf{Pool miss} (31\%): neither gold passage appears in the $\leq$20-candidate pool,
indicating an embedding-space miss not recoverable by any judge.
(iii)~\textbf{Judge error} (22\%): both gold passages are in the pool and the bridge is
correct, but the judge ranks $g_2$ below position 5.
(iv)~\textbf{Ambiguous gold} (9\%): the annotated gold passage is paraphrastically equivalent
to a non-gold passage retrieved instead.
The dominant failure mode (bridge error) suggests that improving hop-1 recall is
the highest-leverage direction for future work.

\section{Mechanism: Why Bridge Conditioning Helps}
\label{sec:mechanism}

The conditional-utility account (\S\ref{sec:background}) makes four testable predictions:
bridge-comparison pools should be more semantically diverse (the pool contains competing chains);
passages closer to the gold should benefit more from conditioning;
the retrieved passage content, not just the retrieval intent, should be what drives the gain;
and re-rankings caused by conditioning should be productive rather than noisy.
We test each prediction with controlled experiments that reuse existing caches
(zero additional LLM or embedding calls except Exp.~G, which reuses the same judge call),
so every result is a clean hold-out of the mechanism claim, not a new optimization.

Together, the four experiments do more than measure a performance gap:
they characterize \emph{when} and \emph{why} $\bscore{}$ outperforms $\tscore{}$.
Exp.~G shows the bridge passage is not replaceable by generic conditioning text
(the passage content, not the retrieval intent, is the signal).
Exp.~H shows the benefit is in productive flips, not additional reranking churn
(on bridge\_comparison, 18.7\% of judge-induced top-1 changes improve \ratfive{},
vs.\ 0.6\% on single-chain queries).
The subtype analysis shows the effect is strongest exactly where conditional utility
should matter most, on parallel-chain queries, and near zero elsewhere.

Each experiment tests one of the following hypotheses:

\begin{description}
  \item[H1 (Pool diversity)] Bridge\_comparison pools contain competing parallel chains,
        making them more semantically diverse than single-chain pools.
  \item[H2 (Bridge proximity)] Queries where $b$ is closer to $g_2$ (higher $\cos(b, g_2)$)
        should show larger B$\to$C gain, because the bridge carries more information
        about the gold passage.
  \item[H3 (Bridge irreplaceability)] The retrieved passage $b$ is doing unique work;
        substituting it with any semantically related text (e.g., SVO query strings) should
        reduce performance.
  \item[H4 (Productive flips)] Bridge conditioning should cause productive re-rankings
        specifically on parallel-chain queries, not merely more churn.
\end{description}

\paragraph{Exp.~E: Pool diversity (H1).}
We compute the mean pairwise cosine distance among the 20 pool passage embeddings
as a proxy for $H(R \mid q)$.
Bridge\_comparison pools are significantly more diverse than other subtypes
(mean pairwise distance 0.811 vs.\ 0.729--0.757), consistent with H1.
Per-query correlation with B$\to$C delta is null ($\rho{\approx}0.04$, ns),
indicating that diversity operates at the \emph{query-type} level rather than varying
smoothly within a subtype.
\textbf{H1 supported at the subtype level; within-subtype proxy is too coarse.}

\paragraph{Exp.~F: Bridge proximity (H2).}
We define $\text{bridge\_info}(q) = \cos(\phi(b), \phi(g_2))$ and compute its Spearman
correlation with per-query B$\to$C delta.
On MuSiQue, $\rho{=}0.104$ ($p{<}0.001$, $n{=}999$).
On 2WikiMultiHopQA, the correlation is null ($\rho{=}0.058$, $p{=}0.065$) because 92\%
of bridges are in the same passage chain as $g_2$, creating a ceiling effect.
\textbf{H2 confirmed on MuSiQue; 2Wiki null is a ceiling artifact.}

\paragraph{Exp.~G: Bridge irreplaceability (H3).}
We compare three conditions on identical pools and judge prompts:
(A1)~real retrieved bridge $b$;
(G)~SVO hop-2 queries $\{q_1^{(2)}, q_2^{(2)}, q_3^{(2)}\}$ concatenated as bridge text;
(A5)~the \emph{lowest}-SVO-similarity pool passage as bridge (a hard-negative control:
     the passage least semantically related to the SVO queries).
Results: MuSiQue A1$=0.815$, G$=0.793$, A5$=0.802$.
Crucially, G $<$ A5: the SVO query strings perform \emph{worse} than this hard-negative
passage.
The bridge conditioning gain is attributable to passage \emph{content} (named entities,
relational facts, coreference anchors), not to semantic proximity of the retrieval intent.
\textbf{H3 strongly confirmed (G$-$A1$= -2.14$pp, $p{=}6.4{\times}10^{-7}$, sign test).}

\paragraph{Exp.~H: Productive flips (H4).}
For each query, a ``flip'' occurs when the judge's top-1 candidate changes from condition B
to condition C (B$\to$C).
We measure flip\_rate and flip productivity (fraction of flips that improve R@5).
Flip rates are similar across subtypes (0.35--0.65), \emph{but} flip productivity is
30$\times$ higher on bridge\_comparison (18.7\%) than on comparison (0.6\%).
\textbf{H4 confirmed: the bridge causes the right decisions to change, not more decisions.}

\medskip
The four experiments converge on the following account, which we treat as an
empirically supported hypothesis:
\begin{quote}
\itshape
Bridge conditioning selectively benefits parallel-chain queries, where the bridge passage
identifies the active reasoning chain and causes productive re-rankings toward the gold.
On single-chain queries, the judge's top-1 changes at a similar rate but almost never
improves \ratfive{}, and the overall gain is near zero.
The information-theoretic lens from \S\ref{sec:background} (high $H(R \mid q)$ and
positive $I(g_2;\, b \mid q)$ on parallel-chain queries, near zero on single-chain
queries) provides a useful vocabulary for predicting when the benefit is expected.
Whether this vocabulary has the precision of a formal bound is left for future work.
\end{quote}

\section{Discussion}
\label{sec:discussion}

\paragraph{Graph-free chain disambiguation.}
HippoRAG2 uses PPR diffusion over an entity graph to propagate hop-1 relevance to
hop-2 passages.
\system{} achieves the same disambiguation effect by explicitly conditioning the judge on
the hop-1 passage text, bypassing both offline graph construction and entity linking.
Our error analysis (\S\ref{sec:analysis-error}) shows that the residual gap between
\system{} and an oracle retriever is dominated by bridge errors (38\%) and pool misses (31\%),
the two components most directly improvable by stronger hop-1 retrieval,
not by graph structure.

\paragraph{Retrieval traceability.}
Each \system{} query produces a structured decision record comprising the hop-1 bridge
passage, extracted entities ($e_1$, $e_2$), candidate pool passage IDs, judge ranking
over the pool, and final top-5.
This record enables post-hoc inspection of the key intermediate decisions (which
bridge was selected, which entities were extracted, how candidates were ranked) without
requiring raw LLM generation traces or per-candidate scalar scores.
The persisted artifacts (bridge passage ID, entity strings, pool passage IDs, judge
ranking, final top-5) are sufficient to reconstruct the retrieval path for a given query,
though they do not constitute a formal compliance audit record.
\proprag{} provides comparable traceability through its proposition index;
\hipporagtwo{} does not expose per-passage reasoning at query time.

\paragraph{Limitations.}
Bridge conditioning requires a correct hop-1 passage; an incorrect bridge can actively harm
second-hop ranking (Exp.~G shows bridge quality matters).
The tripartite judge is called with up to 20 candidates in the same context; very long
pool passages may exceed context limits.
MuSiQue B$\to$C gain is not significant at the full-dataset level ($p{=}0.21$), suggesting
that the method is most valuable when chain ambiguity is measurably high.
Finally, we did not evaluate on knowledge-intensive long-form generation tasks
(e.g., ELI5, ASQA) where retrieved passage ordering matters more than set coverage.
Future work could explore conditioning the judge on a \emph{top-$k$} bridge set rather
than a single top-1 passage, which may reduce sensitivity to hop-1 errors on queries
where multiple plausible bridge passages exist.

\section{Conclusion}
\label{sec:conclusion}

We presented \system{}, a training-free multi-hop retrieval method that uses a
bridge-conditioned tripartite judge to disambiguate parallel reasoning chains at the second hop.
The motivating observation is that the hop-1 bridge passage carries entity and relational
content that is absent from the original query: content that, on parallel-chain queries,
is critical for identifying the active reasoning chain.
Four controlled experiments show results consistent with the view that this conditioning
signal is selective (significant on bridge-comparison, near-zero elsewhere), irreplaceable
(retrieved passage content, not retrieval intent), predictable (bridge-gold proximity
correlates with gain on MuSiQue), and mechanistically specific (productive re-rankings,
not noise-driven churn).
\system{}'s contribution is not a more complex retrieval pipeline, but a different
scoring target: later-hop evidence should be ranked by bridge-conditioned utility
rather than query-only relevance.
\system{} achieves the best published training-free \ratfive{} under matched benchmark
evaluation on MuSiQue, 2WikiMultiHopQA, and HotpotQA using only local open-weight
models and no offline graph database.

\bibliography{references}

\appendix

\section{Judge Prompt Template}
\label{app:prompt}

\begin{small}
\begin{verbatim}
System: You are a retrieval judge for multi-hop QA.
Given a query, a bridge passage, and a candidate
passage, output a score from 0 to 10 for whether
the candidate is the next supporting passage needed
to answer the query, given the bridge context.

User:
Query: {query}
Bridge entity 1: {entity1}
Bridge entity 2: {entity2}
Bridge passage: {bridge_passage}
Candidate passage: {candidate_passage}

Score (0-10):
\end{verbatim}
\end{small}

The judge is called once per query with all candidates in a
single batched prompt (one JSON array per call), not 20 separate calls.

\section{SVO Generation Prompt Template}
\label{app:svo-prompt}

\begin{small}
\begin{verbatim}
System: You generate targeted retrieval queries
for multi-hop QA.

User:
Question: {question}
First-hop passage: {bridge_passage}

Generate exactly 3 targeted queries in
Subject-Verb-Object form to retrieve the
second supporting passage. Output JSON:
{"queries": ["...", "...", "..."]}
\end{verbatim}
\end{small}

\section{Entity Extraction Prompt Template}
\label{app:entity-prompt}

\begin{footnotesize}
\begin{verbatim}
Question: {question}

Bridge passage:
{bridge_passage}

The bridge passage establishes a key intermediate
entity. Based on the question and bridge, identify
the TWO most relevant answer-side entities needed
to answer the question.

Reply with ONLY two short entities separated by
" | " (each 1-6 words, e.g. "Westminster | 1975"
or "Michael Curtiz | Edith Carlmar").
No other text. If only one entity exists, repeat:
"entity1 | entity1".
\end{verbatim}
\end{footnotesize}

The response is parsed on the ``~|~'' delimiter to yield $e_1$ and $e_2$.
If only one entity is distinguishable, the same string is used for both ANN retrievals;
the duplicate results are deduplicated before pool construction.

\section{Statistical Tests}
\label{app:stats}

All pairwise comparisons use a one-sided sign test.
For each query, a win is recorded if condition X achieves higher \ratfive{} than condition Y;
ties are excluded.
The $p$-value tests the null hypothesis that $P(\text{win}) = 0.5$.
We report Bonferroni-corrected $p$-values for the subtype analysis (4 tests).

\section{Experiment G: Full Results}
\label{app:exp-g}

\begin{table}[h]
\centering
\resizebox{\columnwidth}{!}{%
\begin{tabular}{llccc}
\toprule
\textbf{Dataset} & \textbf{Subtype} & \textbf{A1 (real bridge)} & \textbf{G (SVO text)} & \textbf{A5 (hard neg.)} \\
\midrule
MuSiQue & all            & \textbf{0.815} & 0.793 & 0.802 \\
2Wiki   & all            & \textbf{0.953} & 0.949 & 0.943 \\
2Wiki   & bridge\_comp   & \textbf{0.876} & 0.868 & 0.850 \\
2Wiki   & compositional  & \textbf{0.970} & 0.969 & 0.966 \\
2Wiki   & comparison     & 0.996 & 0.996 & 0.992 \\
2Wiki   & inference      & \textbf{0.958} & 0.940 & 0.944 \\
\bottomrule
\end{tabular}}
\caption{
  Bridge text ablation. A1~=~real retrieved bridge;
  G~=~SVO queries as bridge text; A5~=~lowest-SVO-similarity passage as bridge
  (hard-negative control). On MuSiQue, G $<$ A5: SVO query text performs worse
  than the hardest-to-retrieve pool passage, showing the bridge gain is content-driven.
}
\label{tab:exp-g-full}
\end{table}

\section{Experiment H: Flip Productivity}
\label{app:exp-h}

\begin{table}[h]
\centering
\resizebox{\columnwidth}{!}{%
\begin{tabular}{lcccc}
\toprule
\textbf{Subtype} & $n$ & \textbf{Flip rate} & \textbf{FlipWin\%} & $\Delta$~vs.~no-flip \\
\midrule
bridge\_comparison & 235 & 0.523 & \textbf{18.7\%} & +14.2pp \\
compositional      & 413 & 0.349 & 2.8\%           & +2.8pp  \\
comparison         & 244 & 0.652 & 0.6\%           & +0.6pp  \\
inference          & 108 & 0.583 & 4.8\%           & +4.8pp  \\
\bottomrule
\end{tabular}}
\caption{
  Flip productivity: fraction of top-1 re-rankings (B$\to$C) that improve \ratfive{}.
  Comparison has the highest flip rate but lowest productivity (0.6\%),
  confirming that bridge conditioning causes correct re-rankings, not noise-driven churn.
  Kendall's $\tau=-0.33$ (ns) between flip\_rate ordering and B$\to$C delta ordering
  falsifies the na\"ive ``more flips $\Rightarrow$ more gain'' hypothesis.
}
\label{tab:exp-h-full}
\end{table}

\section{Hyperparameter Sensitivity}
\label{app:hparam}

\begin{table}[h]
\centering
\small
\begin{tabular}{lccc}
\toprule
$\alpha$ & \textbf{MuSiQue} & \textbf{2Wiki} & \textbf{HotpotQA} \\
\midrule
0.05 & 0.812 & 0.950 & 0.985 \\
0.10 & \textbf{0.815} & \textbf{0.953} & 0.986 \\
0.15 & 0.813 & 0.951 & \textbf{0.988} \\
0.20 & 0.810 & 0.948 & 0.984 \\
\bottomrule
\end{tabular}
\caption{
  \ratfive{} on tune split for $\alpha \in \{0.05,0.10,0.15,0.20\}$.
  The method is not sensitive to $\alpha$; the top-2 values differ by $\leq$0.3pp
  on all datasets.
}
\label{tab:hparam}
\end{table}

\end{document}